\documentclass[10pt,journal,comsoc]{IEEEtran}

\usepackage{mathtools}
\usepackage[boxruled]{algorithm2e}
\usepackage{blindtext, graphicx}
\usepackage{relsize}
\usepackage{cite}
\usepackage{bm}
\usepackage{changepage}
\usepackage{amssymb}
\usepackage{array}
\usepackage{longtable}
\usepackage[long,nocomma]{optidef}
\usepackage{framed}
\usepackage{subfigure}
\usepackage[table,xcdraw]{xcolor}
\usepackage[hyphens]{url}
\usepackage{hyperref}

\renewcommand{\arraystretch}{1.2}

\begin{document}

\title{Fast and Efficient Bulk Multicasting over Dedicated Inter-Datacenter Networks}

\author{Mohammad~Noormohammadpour,\textsuperscript{1} Cauligi~S.~Raghavendra,\textsuperscript{1} Srikanth Kandula,\textsuperscript{2} Sriram Rao\textsuperscript{3}\\%
\textsuperscript{1}University of Southern California, \textsuperscript{2}Microsoft, \textsuperscript{3}Facebook
\IEEEcompsocitemizethanks{\IEEEcompsocthanksitem A preliminary version of this paper appears in INFOCOM 2018 \cite{quickcast}}%
}



\maketitle

\begin{abstract}
Several organizations have built multiple datacenters connected via dedicated wide area networks over which large inter-datacenter transfers take place. This includes tremendous volumes of bulk multicast traffic generated as a result of data and content replication. Although one can perform these transfers using a single multicast forwarding tree, that can lead to poor performance as the slowest receiver on each tree dictates the completion time for all receivers. Using multiple trees per transfer each connected to a subset of receivers alleviates this concern. The choice of multicast trees also determines the total bandwidth usage. To further improve the performance, bandwidth over dedicated inter-datacenter networks can be carved for different multicast trees over specific time periods to avoid congestion and minimize the average receiver completion times.

In this paper, we break this problem into the three sub-problems of partitioning, tree selection, and rate allocation. We present an algorithm called QuickCast which is computationally fast and allows us to significantly speed up multiple receivers per bulk multicast transfer with control over extra bandwidth consumption. We evaluate QuickCast against a variety of synthetic and real traffic patterns as well as real WAN topologies. Compared to performing bulk multicast transfers as separate unicast transfers, QuickCast achieves up to $3.64\times$ reduction in mean completion times while at the same time using $0.71\times$ the bandwidth. Also, QuickCast allows the top $50\%$ of receivers to complete between $3\times$ to $35\times$ faster on average compared with when a single forwarding multicast tree is used for data delivery.
\end{abstract}


\section{Introduction}
Dedicated inter-datacenter networks connect dozens of geographically dispersed datacenters \cite{b4, facebook-express-backbone, swan-backbone} whose traffic can generally be categorized as either user-generated or internal. User-generated traffic is in the critical path of users' quality of experience and is generated as a result of direct interaction with users. Internal traffic flows across servers that host applications in the back-end and is a product of staging data and content that will later be used to offer services to users. Compared to user-generated traffic, internal traffic is more resilient to scheduling and routing latency and is usually orders of magnitude larger in volume \cite{b4, tempus, facebook-express-backbone}. Internal data transfers over inter-datacenter networks can potentially take a long time to complete, that is, up to hours \cite{tempus}.

A prevalent form of internal traffic is the replication of data and content from one datacenter to multiple other datacenters which accounts for tremendous volumes of traffic \cite{b4,facebook-express-backbone,overlay_hkust}. Examples include the distribution of numerous copies of voluminous configuration files \cite{configurator}, multimedia content served to regional users by CDNs \cite{netflix-replication}, and search index updates \cite{b4}. Such replication generates bulk multicast transfers with a predetermined set of receivers and known transfer volume which are the focus of this paper.

As bandwidth over dedicated inter-datacenter networks is managed by one organization that also operates the datacenters, it is possible to coordinate data transmissions across the end-points to avoid congestion and optimize network-wide performance metrics such as mean or tail completion times of receivers. We focus on minimizing the mean completion times of receivers while performing concurrent bulk multicast transfers assuming that receivers of a transfer can complete at different times. Speeding up several receivers per transfer can translate to improved end-user quality of experience and increased availability. For example, faster replication of video content to regional datacenters enhances average user's experience in social media applications or making a newly trained model available at regional datacenters allows speedier access to new application features for millions of users.

Several recent works focus on improving the performance of unicast transfers over dedicated inter-datacenter networks \cite{b4, swan, tempus, amoeba, owan}. Performing bulk multicast transfers as many separate unicast transfers can lead to excessive bandwidth usage and increase completion times. Although there exists extensive work on multicasting, it is not possible to apply those solutions to our problem as existing research has focused on different goals and considers different constraints. For example, earlier research in multicasting aims at dynamically building and pruning multicast trees as receivers join or leave \cite{ip_multicast}, building multicast overlays that reduce control traffic overhead and improve scalability \cite{nice}, or choosing multicast trees that satisfy a fixed available bandwidth across all edges as requested by applications \cite{online_multicast_bw_guarantees, sdn_multicast}, minimize congestion within datacenters \cite{avalanche, datacast}, reduce data recovery costs assuming some recovery nodes \cite{raera}, or maximize the throughput of a single multicast flow \cite{MPMC_2013, MPMC_2016}. To our knowledge, none of the related research efforts aimed at minimizing the mean completion times of receivers for concurrent bulk multicast transfers while considering the overall bandwidth usage, which is the focus of this work.

\begin{figure}[t]
    \centering
    \includegraphics[width=\columnwidth]{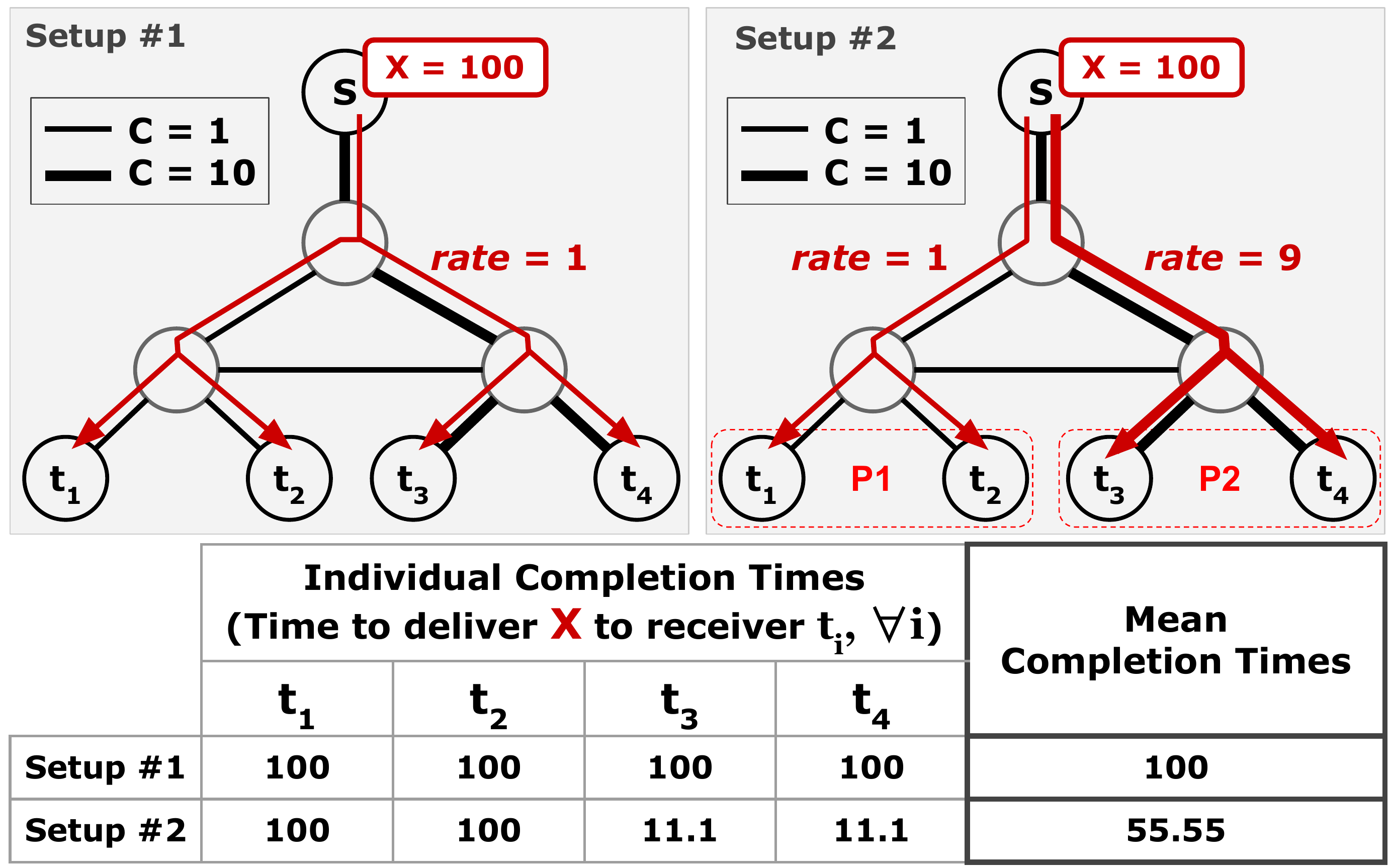}
    \caption{Motivating Example}
    \label{fig:motivating_example}
\end{figure}

\textbf{Motivating Example:} Figure \ref{fig:motivating_example} shows an example of delivering a large object \textbf{X} from source $S$ to destinations $\{t_1,t_2,t_3,t_4\}$ which has a volume of 100 units. We have two types of links with capacities of 1 and 10 units of traffic per time unit. We can use a single multicast tree to connect the sender to all receivers which will allow us to transmit at the bottleneck rate of 1 to all receivers. However, one can group receivers into two partitions of $P1$ and $P2$ and attach each partition with a separate multicast tree. Then we can select transmission rates so that we minimize the mean completion times. In this case, assigning a rate of 1 to the tree attached to $P1$ and a rate of 9 to the tree attached to $P2$ will attain this goal while respecting link capacity over all links (the link attached to $S$ is the bottleneck). As another possibility, we could have assigned a rate of 10 to the tree attached to $P2$, allowing $\{t_3,t_4\}$ to finish in 10 units of time, while suspending the tree attached to $P1$ until time 11. As a result, the tree attached to $P1$ would have started at 11 allowing $\{t_1,t_2\}$ to finish at 110. In this paper, we aim to improve the speed of several receivers per bulk multicast transfer without hurting the completion times of the slow receivers. In computing the completion times, we ignore the propagation and queuing latencies as the focus of this paper is on delivering bulk objects for which the transmission time dominates the propagation or queuing latency along the trees.

We break the bulk multicast transfer routing, and scheduling problem with the objective of minimizing mean completion times of receivers into three sub-problems of the receiver set partitioning, multicast forwarding tree selection per receiver partition, and rate allocation per forwarding tree. We propose QuickCast, which offers an elegant solution to each one of these three sub-problems.\footnote{Compared to \cite{quickcast}, we have extended QuickCast by considering actual WAN topologies with non-uniform link capacity and by eliminating the constraint on the number of partitions. We have performed additional empirical evaluations on multiple tree selection techniques and several rate allocation policies.}

\textit{Receiver Set Partitioning:} As different receivers can have different completion times, a natural way to improve completion times is to partition receivers into multiple sets with each receiver set having a separate tree. This reduces the effect of slow receivers on faster ones. We employ a partitioning technique that groups receivers of every bulk multicast transfer into multiple partitions according to their mutual distance (in hops) on the inter-datacenter graph. With this approach, the partitioning of receivers into any $N > 1$ partitions consumes minimal additional bandwidth on average. We also offer a configuration parameter called the partitioning factor that is used to decide on the right number of partitions that create a balance between receiver completion times improvements and the total bandwidth consumption.

\textit{Forwarding Tree Selection:} To avoid heavily loaded routes, multicast trees should be chosen dynamically per partition according to the receivers in that partition and the distribution of traffic load across network edges. We utilize a computationally efficient approach for forwarding tree selection that connects a sender to a partition of its receivers by assigning weights to edges of the inter-datacenter graph, and using a minimum weight Steiner tree heuristic. We define a weight assignment according to the traffic load scheduled on edges and their capacity and empirically show that this weight assignment offers improved receiver completion times at minimal bandwidth consumption.

\textit{Rate Allocation:} Given the receiver partitions and their forwarding trees, formulating the rate allocation for minimizing mean completion times of receivers leads to a hard problem. We consider the popular scheduling policies of fair sharing, Shortest Remaining Processing Time (SRPT), and First Come First Serve (FCFS). We reason why fair sharing is preferred compared to policies that strictly prioritize transfers (i.e., SRPT, FCFS, etc.) for network throughput maximization when focusing on bulk multicast transfers especially ones with many receivers per transfer. We empirically show that using max-min fairness \cite{max-min-fairness}, which is a form of fair sharing, we can considerably improve the average network throughput which in turn reduces receiver completion times. In QuickCast, we applied max-min fairness for rate allocation across the multicast forwarding trees.

QuickCast assumes a logically centralized setting, communicates with the end-points that transmit traffic for rate limiting, and with the inter-datacenter network elements that perform traffic forwarding for managing multicast forwarding trees. We evaluate QuickCast against a variety of synthetic and real traffic patterns as well as real WAN topologies. Compared to performing bulk multicast transfers as separate unicast transfers, QuickCast achieves up to $3.64\times$ reduction in mean receiver completion times while at the same time using $0.71\times$ the bandwidth. Also, QuickCast allows the top 50\% of receivers to complete between $3\times$ to $35\times$ faster on average compared with when a single forwarding multicast tree is used for data delivery. We also show that on a real WAN topology, fair sharing offers up to $1.5\times$ higher throughput with 16 receivers per bulk multicast transfer compared to other scheduling policies, i.e., SRPT and FCFS.

\section{Related Work} \label{related_work}

\textbf{Internet Multicasting:} 
A large body of general multicasting approaches have been proposed where receivers can join multicast groups anytime to receive required data and multicast trees are incrementally built and pruned as nodes join or leave a multicast session such as IP multicasting \cite{ip_multicast}, TCP-SMO \cite{tcp-smo} and NORM \cite{norm}. These solutions focus on building and maintaining multicast trees, and do not consider link capacity and other ongoing multicast flows while building the trees.

\vspace{0.2em}
\textbf{Multicast Traffic Engineering:}
An interesting work \cite{online_multicast_bw_guarantees} considers the online arrival of multicast requests with a specified bandwidth requirement. The authors provide an elegant solution to find a minimum weight Steiner tree for an arriving request with all edges having the requested available bandwidth. This work assumes a fixed transmission rate per multicast tree, dynamic multicast receivers, and unknown termination time for multicast sessions whereas we consider variable transmission rates over timeslots, fixed multicast receivers, and deem a multicast tree completed when all its receivers download a specific volume of data. MTRSA \cite{sdn_multicast} considers a similar problem to \cite{online_multicast_bw_guarantees} but in an offline scenario where all multicast requests are known beforehand while taking into account the number of available forwarding rules per switch. MPMC \cite{MPMC_2013, MPMC_2016} maximizes the throughput for a single multicast transfer by using multiple parallel multicast trees and coding techniques. None of these works aims to minimize the completion times of receivers while considering the total bandwidth consumption.

\vspace{0.2em}
\textbf{Datacenter Multicasting:}
A variety of solutions have been proposed for minimizing congestion across the intra-datacenter network by selecting multicast trees according to link utilization. Datacast \cite{datacast} sends data over edge-disjoint Steiner trees found by pruning spanning trees over various topologies of FatTree, BCube, and Torus. AvRA \cite{avalanche} focuses on tree and FatTree topologies and builds minimum edge Steiner trees that connect the sender to all receivers as they join. MCTCP \cite{mctcp} reactively schedules flows according to link utilization. These works do not aim at minimizing the completion times of receivers and ignore the total bandwidth consumption.

\vspace{0.2em}
\textbf{Overlay Multicasting:}
With overlay networks, end-hosts can form a multicast forwarding tree in the application layer. RDCM \cite{rdcm} populates backup overlay networks as nodes join and transmits lost packets in a peer-to-peer fashion over them. NICE \cite{nice} creates hierarchical clusters of multicast peers and aims to minimize control traffic overhead. AMMO \cite{AMMO} allows applications to specify performance constraints for selection of multi-metric overlay trees. DC2 \cite{dc2} is a hierarchy-aware group communication technique to minimize cross-hierarchy communication. SplitStream \cite{split-stream} builds forests of multicast trees to distribute load across many machines. BDS \cite{overlay_hkust} generates an application-level multicast overlay network, creates chunks of data, and transmits them in parallel over bottleneck-disjoint overlay paths to the receivers. Due to limited knowledge of underlying physical network topology and condition (e.g., utilization, congestion or even failures), and limited or no control over how the underlying network routes traffic, overlay routing has limited capability in managing the total bandwidth usage and distribution of traffic to minimize completion times of receivers. In case such control and information are provided, for example by using a cross-layer approach, overlay multicasting can be used to realize solutions such as QuickCast.

\vspace{0.2em}
\textbf{Reliable Multicasting:}
Various techniques have been proposed to make multicasting reliable including the use of coding and receiver (negative or positive) acknowledgments. Experiments have shown that using positive ACKs does not lead to ACK implosion for medium scale (sub-thousand) receiver groups \cite{tcp-smo}. TCP-XM \cite{tcp-xm} allows reliable delivery by using a combination of IP multicast and unicast for data delivery and re-transmissions. MCTCP \cite{mctcp} applies standard TCP mechanisms for reliability. Another approach is for receivers to send NAKs upon expiration of some inactivity timer \cite{norm}. NAK suppression has been proposed to address implosion which can be applied by routers \cite{arm}. Forward Error Correction (FEC) has been used to reduce re-transmissions \cite{norm} and improve the completion times \cite{avalanche_code} examples of which include Raptor Codes \cite{raptor} and Tornado Codes \cite{tornado}. These techniques can be applied complementary to QuickCast.

\vspace{0.2em}
\textbf{Multicast Congestion Control:}
Existing approaches track the slowest receiver. PGMCC \cite{pgmcc}, MCTCP \cite{mctcp} and TCP-SMO \cite{tcp-smo} use window-based TCP like congestion control to compete fairly with other flows. NORM \cite{norm} uses an equation-based rate control scheme. With rate allocation and end-host based rate limiting applied over inter-datacenter networks, need for distributed congestion control becomes minimal; however, such techniques can still be used as a backup.

\vspace{0.2em}
\textbf{Other Related Work:}
CastFlow \cite{castflow} precalculates multicast spanning trees which can then be used at request arrival time for fast rule installation. ODPA \cite{odpa} presents algorithms for dynamic adjustment of multicast spanning trees according to specific metrics. BIER \cite{bier} has been recently proposed to improve the scalability and allow frequent dynamic manipulation of multicast forwarding state in the network and can be applied complementary to our solutions in this paper. Peer-to-peer approaches \cite{promise, bittorrent, slurpie} aim to maximize throughput per receiver without considering physical network topology, link capacity, or total network bandwidth consumption. Store-and-Forward (SnF) approaches \cite{netstitcher, mbdt, dtb, mbdt_initial} focus on minimizing transit bandwidth costs which does not apply to dedicated inter-datacenter networks. However, SnF can still be used to improve overall network utilization in the presence of diurnal link utilization patterns, transient bottleneck links, or for application layer multicasting. BDS \cite{bds} uses many parallel overlay paths from a multicast source to its destinations storing and forwarding data from one destination to the next. Application of SnF for bulk multicast transfers considering the physical topology is complementary to our work in this paper and is a future direction. Recent research \cite{ddccast,multicast_deadline,AGE} also consider bulk multicast transfers with deadlines with the objective of maximizing the number of transfers completed before the deadlines. We realize that reducing completion times is a more general objective and for most applications, completing transfers is valuable and required even when it is not possible to meet all the deadlines \cite{tempus}.

\section{Problem Statement and Challenges}
We consider a scenario where bulk multicast transfers arrive at the inter-datacenter network in an online fashion. Every bulk multicast transfer $R$ is specified with a source $S_R$, set of destinations $\pmb{\mathrm{D}}_{R}$, and volume $\mathcal{V}_{R}$ in bytes (unicast and broadcast can be considered as special cases with one receiver or all other nodes as receivers). In general, no form of synchronization is required across receivers of a bulk multicast transfer and therefore, receivers are allowed to complete at different times as long as they all receive the multicast object in whole. Incoming requests are processed as they come by a traffic engineering server that manages the forwarding state of the whole network in a logically centralized manner for installation and eviction of multicast trees. Upon arrival of a request, this server decides on the number of partitions and receivers that are grouped per partition and a multicast tree per partition.

We consider a slotted timeline with a timeslot duration of $\delta$. Periodically, the traffic engineering server computes the transmission rates for all multicast trees at the beginning of every timeslot and dispatches them to senders for rate limiting. This allows for a congestion free network since the rates are computed according to link capacity constraints and other ongoing transfers. To minimize control plane overhead, partitions and forwarding trees are fixed once they are established for an incoming transfer. In this context, the bulk multicast transfer routing and scheduling problem can be formally stated as follows. A summary of our notations is present in Table \ref{table_var}.

\begin{table}
\caption{Definition of variables} \label{table_var}
\begin{center}
\begin{tabular}{ |p{1.7cm}|p{6.2cm}| }
    \hline
    \textbf{Variable} & \textbf{Definition} \\
    \hline
    \hline
    $t_{now}$ & The current timeslot \\
    \hline
    $e$ & A directed edge \\
    \hline
    $C_e$ & Capacity of $e$ \\
    \hline
    $0 \le U_e \le 1$ & Edge $e$'s bandwidth utilization \\
    \hline
    $G(\pmb{\mathrm{V}},\pmb{\mathrm{E}})$ & A directed inter-datacenter network graph \\
    \hline
    $T$ & A directed Steiner tree \\
    \hline
    $\delta$ & Duration of a timeslot \\
    \hline
    $R$ & A bulk multicast transfer request \\
    \hline
    $S_{R}$ & Source datacenter of $R$ \\
    \hline
    $\pmb{\mathrm{D}}_{R}$ & Set$\langle\rangle$ of destinations of request $R$ \\
    \hline
    $\mathcal{V}_{R}$ & Original volume of $R$ \\
    \hline
    $\pmb{\mathrm{P}}_R$ & Set$\langle\rangle$ of partitions of request $R$ \\
    \hline
    $\pmb{\mathrm{P}}$ & Set$\langle\rangle$ of partitions of all transfers in the system \\
    \hline
    $T_{P}$ & Forwarding tree of some partition $P \in \pmb{\mathrm{P}}$ \\
    \hline
    $r_{P}(t)$ & The transmission rate over forwarding tree of some partition $P \in \pmb{\mathrm{P}}$ at timeslot $t$ \\
    \hline
    $\mathcal{V}_{P}^r$ & Residual volume of some partition $P \in \pmb{\mathrm{P}}$. Therefore, $\mathcal{V}_{P}^r \le \mathcal{V}_{R}$ where $P \in \pmb{\mathrm{P}}_R$ \\
    \hline
    $L_e > 0$ & Edge $e$'s total traffic load at time $t_{now}$, i.e., total outstanding bytes scaled by $e$'s inverse capacity \\
    \hline
    $p_f \ge 1$ & A configuration parameter that determines a partitioning cost threshold \\
    \hline
    $N_{max} \ge 1$ & A configuration parameter that determines the maximum number of partitions per transfer \\
    \hline
\end{tabular}
\end{center}
\end{table}

\vspace{0.2em}
\textbf{Problem Statement:} Given an inter-datacenter network $G(\pmb{\mathrm{V}},\pmb{\mathrm{E}})$ with the edge capacity $C_e, \forall e \in \pmb{\mathrm{E}}$ and the set of all partitions $\{P \in \pmb{\mathrm{P}} ~\vert~ \mathcal{V}^{r}_{P} > 0\}$, for a newly arriving bulk multicast transfer $R (S_{R}, \pmb{\mathrm{D}}_{R}, \mathcal{V}_{R})$, the traffic engineering server needs to compute a set of receiver partitions $\pmb{\mathrm{P}}_{R}$ each with one or more receivers, and select a forwarding tree $T_P$ for every partition $P \in \pmb{\mathrm{P}}_{R}$. In addition, per timeslot $t$, the traffic engineering server needs to compute the rates $r_{P}(t), ~\{P \in \pmb{\mathrm{P}} ~\vert~ \mathcal{V}^{r}_{P} > 0\}$. The objective is to minimize the average time for a receiver to complete data reception while keeping the total bandwidth consumption below a certain threshold compared to the minimum possible, i.e., a minimum edge Steiner tree per transfer.

\vspace{0.2em}
\textbf{Challenges:} Both the number of ways to partition receivers into subsets and the number of candidate forwarding trees per subset grow exponentially with the problem size. It is, in general, not clear how partitioning and selection of forwarding trees correlate with both receiver completion times and total bandwidth usage. Even the simple objective of minimizing the total bandwidth usage is a hard problem. Also, assuming known forwarding trees, selecting transmission rates per timeslot per tree for minimization of mean receiver completion times is a hard problem. Finally, this is an online problem with unknown future arrivals which adds to the complexity.

\section{QuickCast}
As stated earlier, we need to address the three sub-problems of receiver set partitioning, tree selection, and rate allocation. Since the partitioning sub-problem uses the tree selection sub-problem, we first discuss tree selection in the following. As the last problem, we will address rate allocation. Since the total bandwidth usage is a function of transfer properties (number of receivers, transfer volume, and the location of sender and receivers) and the network topology, it is highly sophisticated to design a solution that guarantees a limit on the total bandwidth usage. Instead, we aim to reduce the completion times while minimally increasing bandwidth usage.

\subsection{Forwarding Tree Selection}
The tree selection problem states that given a network topology with link capacity knowledge, how to choose a Steiner tree that connects a sender to all of its receivers. The objective is to minimize the completion time of receivers (all receivers on a tree complete at the same time) while minimally increasing the total bandwidth usage. Since the total bandwidth usage is directly proportional to the number of edges on selected trees, we would want to keep trees as small as possible. Reduction in completion times can be achieved by avoiding edges that have a large outstanding traffic load. In general, this would mean selecting potentially larger trees to go around such edges, if necessary. This effect can be accounted for by assigning proper weights to the edges of the inter-datacenter graph and choosing a minimum weight Steiner tree that connects the sender to a partition of receivers for some bulk multicast transfer. The minimum weight Steiner tree is a hard problem for which many heuristics exist.

\vspace{0.2em}
\textbf{Weight Assignment:} Since we focus on reducing the receiver completion times for bulk transfers where transmission time could be orders of magnitude larger than propagation or queuing delay, conventional routing metrics such as end-to-end latency are not effective. Also, we realized that instantaneous link utilization, which has been extensively used for traffic engineering over WAN, lacks stability over longer time periods which makes it hard to infer how it will change in the near future. Therefore, we will use a new metric we refer to as link load $L_{e}, \forall e \in \pmb{\mathrm{E}}$ that is defined in Table \ref{table_var} and can be computed as follows:

\begin{equation}
L_e = \frac{1}{C_e} \sum_{P \in \pmb{\mathrm{P}} \vert e \in T_{P}} \mathcal{V}_{P}^r
\end{equation}

A link's load is the total outstanding volume of traffic allocated on that link (that we know will cross over that link in the future) divided by its capacity. We can compute this value since we know the volume of incoming transfers and the edges that it will be using. A link's load is a measure of how busy it is expected to be shortly. It increases as new transfers are scheduled on a link, and diminishes as traffic flows through it. To select a forwarding tree from a source to a set of receivers, we use an edge weight of $L_{e} + \frac{\mathcal{V}_{R}}{C_{e}}$ and select a minimum weight Steiner tree. The selected tree will most likely exclude any links that are expected to be highly busy. Addition of the second element in the weight (new request's volume divided by capacity) helps select smaller trees in case there is not much load on most edges.

Algorithm \ref{algo_dccast} applies the weight assignment approach mentioned above to select a forwarding tree that balances the traffic load across available trees and finds a minimum weight Steiner tree using the GreedyFLAC heuristic \cite{Watel2014}. In \S \ref{evaluations}, we explore a variety of weights for forwarding tree selection as shown in Table \ref{table_cost} and see that this weight assignment provides consistently close to minimum values for the three performance metrics of mean and tail receiver completion times as well as total bandwidth usage.

\SetAlgoVlined
\SetInd{1.2em}{0.5em}
\begin{algorithm}[t]
\caption{Forwarding Tree Selection Algorithm} \label{algo_dccast}
{
\SetKw{KwBy}{by}
\SetKwProg{ComputeTree}{ComputeTree}{}{}

\vspace{0.4em}
/* Variables defined in Table \ref{table_var} */

\vspace{0.4em}
\KwIn{Request $R$, partition $P \in \pmb{\mathrm{P}}_{R}$, $G(\pmb{\mathrm{V}},\pmb{\mathrm{E}})$, and $L_{e}, \forall e \in \pmb{\mathrm{E}}$ }

\vspace{0.4em}
\KwOut{A forwarding tree (set of edges)}
    
\vspace{0.4em}
\ComputeTree{$\mathrm{(}P,R\mathrm{)}$}{
    
    \vspace{0.4em}
    Assign $W_e = (L_{e} + \frac{{\cal V}_R}{C_e})$, $\forall e \in \pmb{\mathrm{E}}$\;
    
    \vspace{0.4em}
    Find a minimum weight Steiner tree $T_P$ which connects the nodes $\{S_R\} \cup P$\;
    
    \vspace{0.4em}
    $L_e \gets L_e + \frac{{\cal V}_R}{C_e},~\forall e \in T_{P}$\;
    
    \vspace{0.4em}
    \Return{$T_{P}$}\;
}
}
\end{algorithm}

\vspace{0.2em}
\textbf{Worst-case Complexity:} Algorithm \ref{algo_dccast} computes one minimum weight Steiner tree. For a request $R$, the worst-case complexity of algorithm \ref{algo_dccast} is $O(\lvert \pmb{\mathrm{V}} \rvert^3 \lvert \pmb{\mathrm{D}}_R \rvert^2+\lvert \pmb{\mathrm{E}} \rvert)$ given the complexity of GreedyFLAC \cite{Watel2014}.

\subsection{Receiver Set Partitioning} \label{partitioning}
The maximum transmission rate on a tree is that of the link with minimum capacity. To improve bandwidth utilization of inter-datacenter backbone, we can replace a large forwarding tree with multiple smaller trees each connecting the source to a subset of receivers. By partitioning, we isolate some receivers from the bottlenecks allowing them to receive data at a higher rate. We aim to find a set of partitions each with at least one receiver that allows for reducing the average receiver completion times while minimally increasing the bandwidth usage. Bottlenecks may appear either due to competing transfers or differences in link capacity. In the former case, some edges may be shared by multiple trees which lead to lower available bandwidth per tree. Such conditions may arise more frequently under heavy load. In the latter case, differences in link capacity can increase completion times especially in large networks and with many receivers per transfer.

Receiver set partitioning to minimize the impact of bottlenecks and reduce completion times is a sophisticated open problem. It is best if partitions are selected in a way that no additional bottlenecks are created. Also, increasing the number of partitions may in general increase bandwidth consumption (multiple smaller trees may have more edges in total compared to one large tree). Therefore, we need to come up with the right number of partitions and receivers that are grouped per partition. We propose a partitioning approach, called the hierarchical partitioning, that is computationally efficient and uses a partitioning factor to decide on the number of partitions and receivers that are grouped in those partitions.

\vspace{0.2em}
\textbf{Number of Partitions:} Transfers may have a highly varying number of receivers. Generally, the number of partitions should be computed based on the number of receivers, where they are located in the network, and the network topology. Also, using more partitions can lead to the creation of unnecessary bottlenecks due to shared links. We compute the number of partitions per transfer according to the total traffic load on network edges and considering a threshold that limits the cost of additional bandwidth consumption.

\vspace{0.2em}
\textbf{Limitations of Partitioning:} Partitioning, in general, cannot improve tail completion times of transfers as tail is usually driven by physical resource constraints, i.e., low capacity links or links with high contention.

\vspace{0.2em}
\textbf{Hierarchical Partitioning:} We group receivers into partitions according to their mutual distance which is defined as the number of hops on the shortest hop path that connects any two receivers. Hierarchical clustering \cite{clustering_methods} approaches such as agglomerative clustering can be used to compute the groups by initially assuming that every receiver has its partition and then by merging the two closest partitions at each step which generates a hierarchy of partitioning solutions. Each layer of the hierarchy then gives us one possible solution with a given number of partitions.

With this approach, the partitioning of receivers into any $N > 1$ partitions consumes minimal additional bandwidth on average compared to any other partitioning with $N$ partitions. That is because assigning a receiver to any other partition will likely increase the total number of edges needed to connect the source to all receivers; otherwise, that receiver would not have been grouped with the other receivers in its current partition in the first place. There is, however, no guarantee since hierarchical clustering works based on a greedy heuristic.

After building a partitioning hierarchy, the algorithm selects the layer with the maximum number of partitions whose total sum of tree weights stays below a threshold that can be configured as a system parameter. Choosing the maximum partitions allows us to minimize the effect of slow receivers given the threshold, which is a multiple of the weight of a single tree that would connect the sender to all receivers and can be looked at as a bandwidth budget. We call the multiplication coefficient the partitioning factor $p_f$. Algorithm \ref{algo_quick} shows this process in detail. The partitioning factor $p_f$ plays a vital role in the operation of QuickCast as it determines the extra cost we are willing to pay in bandwidth for improved completion times. In general, a $p_f$ greater than one but close to it should allow partitioning to separate very slow receivers from several other nodes. A $p_f$ that is considerably larger than one may generate too many partitions and potentially create many shared links which reduce throughput and additional edges that increase bandwidth usage. If $p_f$ is less than one, a single partition will be used.

\SetAlgoVlined
\SetInd{1.2em}{0.5em}
\begin{algorithm}[t]
\caption{Compute Partitions and Trees} \label{algo_quick}
{
\SetKw{KwBy}{by}
\SetKwProg{ComputePartitionsAndTrees}{ComputePartitionsAndTrees}{}{}

\vspace{0.4em}
/* Variables defined in Table \ref{table_var} */

\vspace{0.4em}
\KwIn{Request $R(S_R,\pmb{\mathrm{D}}_R,\mathcal{V}_R)$, $G(\pmb{\mathrm{V}},\pmb{\mathrm{E}})$, and $L_{e}, \forall e \in \pmb{\mathrm{E}}$ }

\vspace{0.4em}
\KwOut{Pairs of (partition, forwarding tree)}
    
\vspace{0.4em}
\ComputePartitionsAndTrees{$\mathrm{(}R,N_{max}\mathrm{)}$}{

    \vspace{0.4em}
    Assign $W_e = (L_{e} + \frac{{\cal V}_R}{C_e})$, $\forall e \in \pmb{\mathrm{E}}$\;
    
    \vspace{0.4em}
    Find the minimum weight Steiner tree $T_R$ which connects the nodes $\{S_R\} \cup \pmb{\mathrm{D}}_R$ and its weight $W_{T_{R}}$\;
    
    \vspace{0.4em}
    \ForEach{$(\alpha,\beta) \in \pmb{\mathrm{D}}_R,~\alpha \neq \beta$}{
        \vspace{0.4em}
        $\mathrm{DIST}_{\alpha,\beta} \gets$ number of edges on the minimum hop path from $\alpha$ to $\beta$\;
    }
    
    \vspace{0.4em}
    Compute agglomerative clustering hierarchy for $\pmb{\mathrm{D}}_R$ using average linkage and distance $\mathrm{DIST}$ which will have $l$ clusters at layer $1 \le l \le \lvert \pmb{\mathrm{D}}_R \rvert$\;
    
    \vspace{0.4em}
    \For{$l = \min(N_{max}, \lvert \pmb{\mathrm{D}}_R \rvert)$ \KwTo $2$ \KwBy $-1$}{
    
        \vspace{0.4em}
        $\pmb{\mathrm{P}}_{l} \gets$ set of clusters at layer $l$ of agglomerative hierarchy, each cluster forms a partition\;
        
        \vspace{0.4em}
        \ForEach{$P \in \pmb{\mathrm{P}}_{l}$}{
            \vspace{0.4em}
            Find the minimum weight Steiner tree $T_{P}$ which connects the nodes $\{S_R\} \cup P$\;
        }
        
        \vspace{0.4em}
        \If{$\sum_{P \in \pmb{\mathrm{P}}_{l}} W_{T_{P}} \le p_f \times W_{T_{R}}$}{
            
            \vspace{0.4em}
            \ForEach{$P \in \pmb{\mathrm{P}}_{l}$}{
                \vspace{0.4em}
                $T_{P} \gets$ \textbf{ComputeTree}~($P$,$R$)\;
            }
            
            \vspace{0.4em}
            \Return{$(P,~T_{P}),~\forall P \in \pmb{\mathrm{P}}_{l}$}\;
        }
    }
    
    \vspace{0.4em}
    $L_e \gets L_e + {\cal V}_R,~\forall e \in T_{R}$\;
    
    \vspace{0.4em}
    \Return{$(\pmb{\mathrm{D}}_R,~T_{R})$}\;
}
}
\end{algorithm}

\vspace{0.2em}
\textbf{Worst-case Complexity:} Algorithm \ref{algo_quick} performs multiple calls to the GreedyFLAC algorithm \cite{Watel2014}. It also uses the hierarchical clustering with average linkage which has a worst-case complexity of $O(\lvert \pmb{\mathrm{D}}_R \rvert^3)$. To compute the pair-wise distances of receivers, we can use breadth first search with has a complexity of $O(\lvert \pmb{\mathrm{V}} \rvert+\lvert \pmb{\mathrm{E}} \rvert)$. Worst-case complexity of Algorithm \ref{algo_quick} is $O((\lvert \pmb{\mathrm{V}} \rvert^3+\lvert \pmb{\mathrm{E}} \rvert) \lvert \pmb{\mathrm{D}}_R \rvert^2+\lvert \pmb{\mathrm{D}}_R \rvert^3)$.

\subsection{Rate Allocation} \label{rate-allocation}
To compute the transmission rates per tree per timeslot, one can formulate an optimization problem with the capacity and demand constraints, and consider minimizing the mean receiver completion times as the objective. This is, however, a hard problem and can be modeled using mixed-integer programming by assuming a binary variable per timeslot per tree that shows whether that tree has completed by that timeslot. One can come up with approximation algorithms to this problem which is considered part of the future work.

In this paper, we consider the three popular scheduling policies of FCFS, SRPT, and fair sharing according to max-min fairness \cite{max-min-fairness} which have been extensively used for network scheduling. These policies can be applied independently of partitioning and forwarding tree selection techniques. Each one of these three policies has its unique features. FCFS and SRPT both prioritize transfers; the former according to arrival times and the latter according to transfer volumes and so obtain a meager fairness score \cite{fairness_theory}. SRPT has been extensively used for minimizing flow completion times within datacenters \cite{pfabric, pias, epn}. Strictly prioritizing transfers over forwarding trees (as done by SRPT and FCFS), however, can lead to low overall link utilization and increased completion times, especially when trees are large. This might happen due to bandwidth contention on shared edges which can prevent some transfers from making progress. Fair sharing allows all transfers to make progress which mitigates such contention enabling concurrent multicast transfers to all make progress. In \S \ref{eval-rate-alloc}, we empirically compare the performance of these scheduling policies and show that fair sharing based on max-min fairness can significantly outperform both FCFS and SRPT in average network throughput especially with a larger number of receivers per tree. As a result, we will use QuickCast along with the fair sharing policy based on max-min fairness.

The traffic engineering server periodically computes the transmission rates per multicast tree every timeslot to maximize utilization and cope with inaccurate inter-datacenter link capacity measurements, imprecise rate limiting, and dropped packets due to corruption. To account for inaccurate rate limiting, dropped packets and link capacity estimation errors, which all can lead to a difference between the actual volume of data delivered and the number of bytes transmitted, we propose that senders keep track of actual data delivered to their receivers per forwarding tree. At the end of every timeslot, every sender reports to the traffic engineering server how much data it was able to deliver allowing it to compute rates accordingly for the timeslot that follows. Newly arriving transfers will be assigned rates starting the next timeslot.

\section{Evaluation} \label{evaluations}
We considered various topologies and transfer size distributions as shown in Tables \ref{table_topology} and \ref{table_traffic}. Also, for Algorithm \ref{algo_quick}, unless otherwise stated, we used $p_f = 1.1$ which limits the overall bandwidth usage while offering significant gains. In the following subsections, we first evaluated a variety of weight assignments for multicast tree selection considering receiver completion times and bandwidth usage. We showed that the weight proposed in Algorithm \ref{algo_dccast} offers close to minimum completion times with minimal extra bandwidth consumption. Next, we evaluated the proposed partitioning technique and considered two cases of $N_{max}=2$ (as used in \cite{quickcast}) and $N_{max}=\lvert \pmb{\mathrm{D}}_R \rvert$. We measured the performance of QuickCast while varying the number of receivers and showed that it offers consistent gains. We also measured the speedup observed by different receivers ranked by their speed per multicast transfer, and the effect of partitioning factor $p_f$ on the gains in completion times as well as bandwidth usage. In addition, we evaluated the effect of different scheduling policies on average network throughput and showed that with increasing number of multicast receivers, fair sharing offers higher throughput compared to both FCFS and SRPT. Finally, we showed that QuickCast is computationally fast by measuring its running time and that the maximum number of group table forwarding entries it uses across all switches is only a fraction of what is usually available in a physical switch across the several considered scenarios.

\vspace{0.2em}
\textbf{Network Topologies:}
Table \ref{table_topology} shows the list of topologies we considered. These topologies provide capacity information for all links which range from 45 Mbps to 10 Gbps. We normalized all link capacities dividing them by the maximum link capacity. We also assumed all bidirectional links with equal capacity in either direction.

\begin{table}
\caption{Various topologies used in evaluation} \label{table_topology}
\begin{center}
\begin{tabular}{ |p{1.8cm}|p{6.0cm}| }
    \hline
    \textbf{Name} & \textbf{Description} \\
    \hline
    \hline
    ANS \cite{ans} & A medium-sized backbone and transit network that spans across the United States with $18$ nodes and $25$ links. All links have equal capacity of 45 Mbps. \\
    \hline
    GEANT \cite{geant} & A large-sized backbone and transit network that spans across the Europe with $34$ nodes and $52$ links. Link capacity ranges from 45 Mbps to 10 Gbps. \\
    \hline
    UNINETT \cite{uninett} & A large-sized backbone that spans across Norway with $69$ nodes and $98$ links. Most links have a capacity of 1, 2.5 or 10 Gbps. \\
    \hline
\end{tabular}
\end{center}
\end{table}

\vspace{0.2em}
\textbf{Traffic Patterns:}
Table \ref{table_traffic} shows the considered distributions for transfer volumes. Transfer arrival followed a Poisson distribution with rate $\lambda$. We considered no units for time or bandwidth. For all simulations, we assumed a timeslot length of $\delta = 1.0$. For Pareto distribution, we considered a minimum transfer volume equal to that of $2$ full timeslots and limited maximum transfer volume to that of $2000$ full timeslots. Unless otherwise stated, we considered an average demand equal to volume of $20$ full timeslots per transfer for all traffic distributions (we fixed the mean values of all distributions to the same value). Per simulation instance, we assumed equal number of transfers per sender and for every transfer, we selected the receivers from all existing nodes according to the uniform distribution (with equal probability from all nodes).

\vspace{0.2em}
\textbf{Assumptions:} We focused on computing gains and assumed accurate knowledge of inter-datacenter link capacity, and precise rate control at the end-points which together lead to a congestion free network. We also assumed no dropped packets due to corruption or errors, and no link failures.

\vspace{0.2em}
\textbf{Simulation Setup:} We developed a simulator in Java (JDK 8). We performed all simulations on one machine (Core i7-6700 and 24 GB of RAM). We used the Java implementation of GreedyFLAC \cite{DSTAlgoEvaluation} for minimum weight Steiner trees.

\begin{table}
\caption{Transfer size distributions (parameters in \S \ref{evaluations})} \label{table_traffic}
\begin{center}
\begin{tabular}{ |p{3cm}|p{4.8cm}| }
    \hline
    \textbf{Name} & \textbf{Description} \\
    \hline
    \hline
    Light-tailed & Based on Exponential distribution. \\
    \hline
    Heavy-tailed & Based on Pareto distribution. \\
    \hline
    Facebook \textit{Cache-Follower} & Generated by cache applications over Facebook inter-datacenter WAN \cite{social_inside}. \\
    \hline
    Facebook \textit{Hadoop} & Generated by geo-distributed analytics over Facebook inter-datacenter WAN \cite{social_inside}. \\
    \hline
\end{tabular}
\end{center}
\end{table}

\subsection{Weight Assignment Techniques for Tree Selection}
We empirically evaluate and analyze several weights for selection of forwarding trees. Table \ref{table_cost} lists the weight assignment approaches considered for tree selection (please see Table \ref{table_var} for definition of variables). We considered three edge weight metrics of utilization (i.e., the fraction of a link's bandwidth currently in use), load (i.e., the total volume of traffic that an edge will carry starting current time), and load plus the volume of the newly arriving transfer request. We also considered the weight of a tree to be either the weight of its edge with maximum weight or the sum of weights of its edges. An exponential weight is used to approximate selection of trees with minimum highest weight, similar to the approach used in \cite{tempus}. The benefit of the weight \#6 over \#5 is that in case there is no load or minimal load on some edges, selecting the minimum weight tree will lead to minimum edge trees that reduce bandwidth usage. Also, with this approach, we tend to avoid large trees for large transfers which helps further reduce bandwidth usage.

\begin{table}[b]
\caption{Various weights for tree selection for incoming request $R$} \label{table_cost}
\begin{center}
\renewcommand{\arraystretch}{1.5}
\begin{tabular}{ |p{0.2cm}|p{2.3cm}|p{5cm}| }
    \hline
    \textbf{\#} & $W_e, \forall e \in \pmb{\mathrm{E}}$ & \textbf{Properties of Selected Trees} \\
    \hline
    \hline
    1 & $1.0$ & A fixed minimum edge Steiner tree \\
    \hline
    2 & $\exp(U_e)$ & Minimum highest utilization over edges \\
    \hline
    3 & $\exp(L_e)$ & Minimum highest load over edges \\
    \hline
    4 & $U_e$ & Minimum sum of utilization over edges \\
    \hline
    5 & $L_e$ & Minimum sum of load over edges \\
    \hline
    6 & $L_e+\frac{\mathcal{V}_{R}}{C_e}$ & Minimum final sum of load over edges \\[0.5mm]
    \hline
    7 & $1.0+\frac{\exp(U_e)} {\sum_{e \in \pmb{\mathrm{E}}} \exp(U_e)}$ & Minimum edges, min-max utilization  \\[1mm]
    \hline
    8 & $1.0+\frac{\exp(L_e)} {\sum_{e \in \pmb{\mathrm{E}}} \exp(L_e)}$ & Minimum edges, min-max load \\[1mm]
    \hline
    9 & $1.0+\frac{U_e} {\sum_{e \in \pmb{\mathrm{E}}} U_e}$ & Minimum edges, min-sum of utilization \\[1mm]
    \hline
    10 & $1.0+\frac{L_e} {\sum_{e \in \pmb{\mathrm{E}}} L_e}$ & Minimum edges, min-sum of load \\[1mm]
    \hline
\end{tabular}
\end{center}
\end{table}

\begin{figure}[t]
\centering
\resizebox{\columnwidth}{!}{
\begin{tabular}{c|c|c|c|c|c|c|c|c|c|c|c|c|}
\cline{2-13}
\multicolumn{1}{l|}{} & \multicolumn{12}{c|}{Mean Receiver Completion Times} \\ \cline{2-13}
\multicolumn{1}{l|}{} & \multicolumn{6}{c|}{ANS} & \multicolumn{6}{c|}{GEANT} \\ \cline{2-13} 
\multicolumn{1}{l|}{} & \multicolumn{3}{c|}{Light-tailed} & \multicolumn{3}{c|}{Heavy-tailed} & \multicolumn{3}{c|}{Light-tailed} & \multicolumn{3}{c|}{Heavy-tailed} \\ \hline 
\multicolumn{1}{|c|}{\cellcolor[HTML]{EFEFEF}\#} & $\mathcal{F}$ & $\mathcal{S}$ & $\mathcal{M}$ & $\mathcal{F}$ & $\mathcal{S}$ & $\mathcal{M}$ & $\mathcal{F}$ & $\mathcal{S}$ & $\mathcal{M}$ & $\mathcal{F}$ & $\mathcal{S}$ & $\mathcal{M}$ \\ \hline
\multicolumn{1}{|c|}{\cellcolor[HTML]{EFEFEF}1}  & \cellcolor[HTML]{34FF34}10- & \cellcolor[HTML]{34FF34}10- & \cellcolor[HTML]{F8FF00}20- & \cellcolor[HTML]{F8FF00}20- & \cellcolor[HTML]{34FF34}10- & \cellcolor[HTML]{F8FF00}20- & \cellcolor[HTML]{F56B00}50- & \cellcolor[HTML]{F8A102}40- & \cellcolor[HTML]{FE0000}50+ & \cellcolor[HTML]{FE0000}50+ & \cellcolor[HTML]{F8A102}40- & \cellcolor[HTML]{F8A102}40- \\ \hline
\multicolumn{1}{|c|}{\cellcolor[HTML]{EFEFEF}2}  & \cellcolor[HTML]{F8FF00}20- & \cellcolor[HTML]{F8FF00}20- & \cellcolor[HTML]{34FF34}10- & \cellcolor[HTML]{F8FF00}20- & \cellcolor[HTML]{FFCC67}30- & \cellcolor[HTML]{34FF34}10- & \cellcolor[HTML]{34FF34}10- & \cellcolor[HTML]{F8FF00}20- & \cellcolor[HTML]{34FF34}10- & \cellcolor[HTML]{F8FF00}20- & \cellcolor[HTML]{34FF34}10- & \cellcolor[HTML]{34FF34}10- \\ \hline
\multicolumn{1}{|c|}{\cellcolor[HTML]{EFEFEF}3}  & \cellcolor[HTML]{F8FF00}20- & \cellcolor[HTML]{F8FF00}20- & \cellcolor[HTML]{34FF34}10- & \cellcolor[HTML]{F8FF00}20- & \cellcolor[HTML]{F56B00}50- & \cellcolor[HTML]{FFCC67}30- & \cellcolor[HTML]{34FF34}10- & \cellcolor[HTML]{34FF34}10- & \cellcolor[HTML]{34FF34}10- & \cellcolor[HTML]{34FF34}10- & \cellcolor[HTML]{34FF34}10- & \cellcolor[HTML]{34FF34}10- \\ \hline
\multicolumn{1}{|c|}{\cellcolor[HTML]{EFEFEF}4}  & \cellcolor[HTML]{F8A102}40- & \cellcolor[HTML]{F8A102}40- & \cellcolor[HTML]{34FF34}10- & \cellcolor[HTML]{F8A102}40- & \cellcolor[HTML]{F8A102}40- & \cellcolor[HTML]{34FF34}10- & \cellcolor[HTML]{F8FF00}20- & \cellcolor[HTML]{FFCC67}30- & \cellcolor[HTML]{34FF34}10- & \cellcolor[HTML]{F8FF00}20- & \cellcolor[HTML]{34FF34}10- & \cellcolor[HTML]{34FF34}10- \\ \hline
\multicolumn{1}{|c|}{\cellcolor[HTML]{EFEFEF}5}  & \cellcolor[HTML]{34FF34}10- & \cellcolor[HTML]{34FF34}10- & \cellcolor[HTML]{34FF34}10- & \cellcolor[HTML]{34FF34}10- & \cellcolor[HTML]{F8FF00}20- & \cellcolor[HTML]{F8FF00}20- & \cellcolor[HTML]{34FF34}10- & \cellcolor[HTML]{34FF34}10- & \cellcolor[HTML]{34FF34}10- & \cellcolor[HTML]{34FF34}10- & \cellcolor[HTML]{34FF34}10- & \cellcolor[HTML]{34FF34}10- \\ \hline
\multicolumn{1}{|c|}{\cellcolor[HTML]{EFEFEF}\textbf{6}}  & \cellcolor[HTML]{34FF34}10- & \cellcolor[HTML]{34FF34}10- & \cellcolor[HTML]{34FF34}10- & \cellcolor[HTML]{34FF34}10- & \cellcolor[HTML]{F8FF00}20- & \cellcolor[HTML]{F8FF00}20- & \cellcolor[HTML]{34FF34}10- & \cellcolor[HTML]{34FF34}10- & \cellcolor[HTML]{34FF34}10- & \cellcolor[HTML]{34FF34}10- & \cellcolor[HTML]{34FF34}10- & \cellcolor[HTML]{34FF34}10- \\ \hline
\multicolumn{1}{|c|}{\cellcolor[HTML]{EFEFEF}7}  & \cellcolor[HTML]{34FF34}10- & \cellcolor[HTML]{34FF34}10- & \cellcolor[HTML]{34FF34}10- & \cellcolor[HTML]{34FF34}10- & \cellcolor[HTML]{34FF34}10- & \cellcolor[HTML]{34FF34}10- & \cellcolor[HTML]{F8A102}40- & \cellcolor[HTML]{FFCC67}30- & \cellcolor[HTML]{FFCC67}30- & \cellcolor[HTML]{F8A102}40- & \cellcolor[HTML]{FFCC67}30- & \cellcolor[HTML]{F8FF00}20- \\ \hline
\multicolumn{1}{|c|}{\cellcolor[HTML]{EFEFEF}8}  & \cellcolor[HTML]{34FF34}10- & \cellcolor[HTML]{34FF34}10- & \cellcolor[HTML]{34FF34}10- & \cellcolor[HTML]{34FF34}10- & \cellcolor[HTML]{34FF34}10- & \cellcolor[HTML]{34FF34}10- & \cellcolor[HTML]{F56B00}50- & \cellcolor[HTML]{F8A102}40- & \cellcolor[HTML]{FE0000}50+ & \cellcolor[HTML]{FE0000}50+ & \cellcolor[HTML]{F8A102}40- & \cellcolor[HTML]{F8A102}40- \\ \hline
\multicolumn{1}{|c|}{\cellcolor[HTML]{EFEFEF}9}  & \cellcolor[HTML]{34FF34}10- & \cellcolor[HTML]{34FF34}10- & \cellcolor[HTML]{34FF34}10- & \cellcolor[HTML]{34FF34}10- & \cellcolor[HTML]{34FF34}10- & \cellcolor[HTML]{34FF34}10- & \cellcolor[HTML]{F8A102}40- & \cellcolor[HTML]{F8A102}40- & \cellcolor[HTML]{FFCC67}30- & \cellcolor[HTML]{F8A102}40- & \cellcolor[HTML]{FFCC67}30- & \cellcolor[HTML]{FFCC67}30- \\ \hline
\multicolumn{1}{|c|}{\cellcolor[HTML]{EFEFEF}10} & \cellcolor[HTML]{34FF34}10- & \cellcolor[HTML]{34FF34}10- & \cellcolor[HTML]{34FF34}10- & \cellcolor[HTML]{34FF34}10- & \cellcolor[HTML]{34FF34}10- & \cellcolor[HTML]{34FF34}10- & \cellcolor[HTML]{FE0000}50+ & \cellcolor[HTML]{FE0000}50+ & \cellcolor[HTML]{FE0000}50+ & \cellcolor[HTML]{FE0000}50+ & \cellcolor[HTML]{FE0000}50+ & \cellcolor[HTML]{F56B00}50- \\ \hline
\end{tabular}
}\vspace{0.2em}
\\
\resizebox{\columnwidth}{!}{
\begin{tabular}{c|c|c|c|c|c|c|c|c|c|c|c|c|}
\cline{2-13}
\multicolumn{1}{l|}{} & \multicolumn{12}{c|}{Tail Receiver Completion Times} \\ \cline{2-13}
\multicolumn{1}{l|}{} & \multicolumn{6}{c|}{ANS} & \multicolumn{6}{c|}{GEANT} \\ \cline{2-13}
\multicolumn{1}{l|}{} & \multicolumn{3}{c|}{Light-tailed} & \multicolumn{3}{c|}{Heavy-tailed} & \multicolumn{3}{c|}{Light-tailed} & \multicolumn{3}{c|}{Heavy-tailed} \\ \hline
\multicolumn{1}{|c|}{\cellcolor[HTML]{EFEFEF}\#} & $\mathcal{F}$ & $\mathcal{S}$ & $\mathcal{M}$ & $\mathcal{F}$ & $\mathcal{S}$ & $\mathcal{M}$ & $\mathcal{F}$ & $\mathcal{S}$ & $\mathcal{M}$ & $\mathcal{F}$ & $\mathcal{S}$ & $\mathcal{M}$ \\ \hline
\multicolumn{1}{|c|}{\cellcolor[HTML]{EFEFEF}1}  & \cellcolor[HTML]{F8FF00}20- & \cellcolor[HTML]{F8FF00}20- & \cellcolor[HTML]{FFCC67}30- & \cellcolor[HTML]{F8FF00}20- & \cellcolor[HTML]{34FF34}10- & \cellcolor[HTML]{F8FF00}20- & \cellcolor[HTML]{F56B00}50- & \cellcolor[HTML]{F56B00}50- & \cellcolor[HTML]{FE0000}50+ & \cellcolor[HTML]{FE0000}50+ & \cellcolor[HTML]{FE0000}50+ & \cellcolor[HTML]{F56B00}50- \\ \hline
\multicolumn{1}{|c|}{\cellcolor[HTML]{EFEFEF}2}  & \cellcolor[HTML]{FFCC67}30- & \cellcolor[HTML]{F8FF00}20- & \cellcolor[HTML]{F8FF00}20- & \cellcolor[HTML]{FFCC67}30- & \cellcolor[HTML]{FFCC67}30- & \cellcolor[HTML]{F8FF00}20- & \cellcolor[HTML]{F8FF00}20- & \cellcolor[HTML]{FFCC67}30- & \cellcolor[HTML]{F8FF00}20- & \cellcolor[HTML]{FFCC67}30- & \cellcolor[HTML]{F8FF00}20- & \cellcolor[HTML]{34FF34}10- \\ \hline
\multicolumn{1}{|c|}{\cellcolor[HTML]{EFEFEF}3}  & \cellcolor[HTML]{F8FF00}20- & \cellcolor[HTML]{F8FF00}20- & \cellcolor[HTML]{34FF34}10- & \cellcolor[HTML]{34FF34}10- & \cellcolor[HTML]{34FF34}10- & \cellcolor[HTML]{34FF34}10- & \cellcolor[HTML]{34FF34}10- & \cellcolor[HTML]{34FF34}10- & \cellcolor[HTML]{34FF34}10- & \cellcolor[HTML]{34FF34}10- & \cellcolor[HTML]{34FF34}10- & \cellcolor[HTML]{34FF34}10- \\ \hline
\multicolumn{1}{|c|}{\cellcolor[HTML]{EFEFEF}4}  & \cellcolor[HTML]{F8A102}40- & \cellcolor[HTML]{F8A102}40- & \cellcolor[HTML]{34FF34}10- & \cellcolor[HTML]{FFCC67}30- & \cellcolor[HTML]{FFCC67}30- & \cellcolor[HTML]{34FF34}10- & \cellcolor[HTML]{FFCC67}30- & \cellcolor[HTML]{FFCC67}30- & \cellcolor[HTML]{F8FF00}20- & \cellcolor[HTML]{F8FF00}20- & \cellcolor[HTML]{F8FF00}20- & \cellcolor[HTML]{34FF34}10- \\ \hline
\multicolumn{1}{|c|}{\cellcolor[HTML]{EFEFEF}5}  & \cellcolor[HTML]{34FF34}10- & \cellcolor[HTML]{34FF34}10- & \cellcolor[HTML]{34FF34}10- & \cellcolor[HTML]{34FF34}10- & \cellcolor[HTML]{34FF34}10- & \cellcolor[HTML]{34FF34}10- & \cellcolor[HTML]{34FF34}10- & \cellcolor[HTML]{34FF34}10- & \cellcolor[HTML]{34FF34}10- & \cellcolor[HTML]{34FF34}10- & \cellcolor[HTML]{34FF34}10- & \cellcolor[HTML]{34FF34}10- \\ \hline
\multicolumn{1}{|c|}{\cellcolor[HTML]{EFEFEF}\textbf{6}}  & \cellcolor[HTML]{34FF34}10- & \cellcolor[HTML]{34FF34}10- & \cellcolor[HTML]{34FF34}10- & \cellcolor[HTML]{34FF34}10- & \cellcolor[HTML]{34FF34}10- & \cellcolor[HTML]{34FF34}10- & \cellcolor[HTML]{34FF34}10- & \cellcolor[HTML]{34FF34}10- & \cellcolor[HTML]{34FF34}10- & \cellcolor[HTML]{34FF34}10- & \cellcolor[HTML]{34FF34}10- & \cellcolor[HTML]{34FF34}10- \\ \hline
\multicolumn{1}{|c|}{\cellcolor[HTML]{EFEFEF}7}  & \cellcolor[HTML]{F8FF00}20- & \cellcolor[HTML]{34FF34}10- & \cellcolor[HTML]{F8FF00}20- & \cellcolor[HTML]{F8FF00}20- & \cellcolor[HTML]{34FF34}10- & \cellcolor[HTML]{34FF34}10- & \cellcolor[HTML]{F8A102}40- & \cellcolor[HTML]{FFCC67}30- & \cellcolor[HTML]{F8A102}40- & \cellcolor[HTML]{F56B00}50- & \cellcolor[HTML]{F8A102}40- & \cellcolor[HTML]{FE0000}50+ \\ \hline
\multicolumn{1}{|c|}{\cellcolor[HTML]{EFEFEF}8}  & \cellcolor[HTML]{34FF34}10- & \cellcolor[HTML]{34FF34}10- & \cellcolor[HTML]{34FF34}10- & \cellcolor[HTML]{34FF34}10- & \cellcolor[HTML]{34FF34}10- & \cellcolor[HTML]{34FF34}10- & \cellcolor[HTML]{F56B00}50- & \cellcolor[HTML]{F56B00}50- & \cellcolor[HTML]{FE0000}50+ & \cellcolor[HTML]{FE0000}50+ & \cellcolor[HTML]{FE0000}50+ & \cellcolor[HTML]{F56B00}50- \\ \hline
\multicolumn{1}{|c|}{\cellcolor[HTML]{EFEFEF}9}  & \cellcolor[HTML]{34FF34}10- & \cellcolor[HTML]{F8FF00}20- & \cellcolor[HTML]{F8FF00}20- & \cellcolor[HTML]{F8FF00}20- & \cellcolor[HTML]{34FF34}10- & \cellcolor[HTML]{34FF34}10- & \cellcolor[HTML]{FFCC67}30- & \cellcolor[HTML]{FFCC67}30- & \cellcolor[HTML]{F8A102}40- & \cellcolor[HTML]{F8A102}40- & \cellcolor[HTML]{FFCC67}30- & \cellcolor[HTML]{FE0000}50+ \\ \hline
\multicolumn{1}{|c|}{\cellcolor[HTML]{EFEFEF}10} & \cellcolor[HTML]{34FF34}10- & \cellcolor[HTML]{34FF34}10- & \cellcolor[HTML]{34FF34}10- & \cellcolor[HTML]{34FF34}10- & \cellcolor[HTML]{34FF34}10- & \cellcolor[HTML]{34FF34}10- & \cellcolor[HTML]{F8A102}40- & \cellcolor[HTML]{FE0000}50+ & \cellcolor[HTML]{F56B00}50- & \cellcolor[HTML]{F8A102}40- & \cellcolor[HTML]{FE0000}50+ & \cellcolor[HTML]{F56B00}50- \\ \hline
\end{tabular}
}\vspace{0.2em}
\\
\resizebox{\columnwidth}{!}{
\begin{tabular}{c|c|c|c|c|c|c|c|c|c|c|c|c|}
\cline{2-13}
\multicolumn{1}{l|}{} & \multicolumn{12}{c|}{Total Bandwidth Used} \\ \cline{2-13}
\multicolumn{1}{l|}{} & \multicolumn{6}{c|}{ANS} & \multicolumn{6}{c|}{GEANT} \\ \cline{2-13}
\multicolumn{1}{l|}{} & \multicolumn{3}{c|}{Light-tailed} & \multicolumn{3}{c|}{Heavy-tailed} & \multicolumn{3}{c|}{Light-tailed} & \multicolumn{3}{c|}{Heavy-tailed} \\ \hline
\multicolumn{1}{|c|}{\cellcolor[HTML]{EFEFEF}\#} & $\mathcal{F}$ & $\mathcal{S}$ & $\mathcal{M}$ & $\mathcal{F}$ & $\mathcal{S}$ & $\mathcal{M}$ & $\mathcal{F}$ & $\mathcal{S}$ & $\mathcal{M}$ & $\mathcal{F}$ & $\mathcal{S}$ & $\mathcal{M}$ \\ \hline
\multicolumn{1}{|c|}{\cellcolor[HTML]{EFEFEF}1}  & \cellcolor[HTML]{34FF34}10- & \cellcolor[HTML]{34FF34}10- & \cellcolor[HTML]{34FF34}10- & \cellcolor[HTML]{34FF34}10- & \cellcolor[HTML]{34FF34}10- & \cellcolor[HTML]{34FF34}10- & \cellcolor[HTML]{34FF34}10- & \cellcolor[HTML]{34FF34}10- & \cellcolor[HTML]{34FF34}10- & \cellcolor[HTML]{34FF34}10- & \cellcolor[HTML]{34FF34}10- & \cellcolor[HTML]{34FF34}10- \\ \hline
\multicolumn{1}{|c|}{\cellcolor[HTML]{EFEFEF}2}  & \cellcolor[HTML]{F8FF00}20- & \cellcolor[HTML]{F8FF00}20- & \cellcolor[HTML]{F8FF00}20- & \cellcolor[HTML]{F8FF00}20- & \cellcolor[HTML]{F8FF00}20- & \cellcolor[HTML]{F8FF00}20- & \cellcolor[HTML]{F8A102}40- & \cellcolor[HTML]{F8A102}40- & \cellcolor[HTML]{F8A102}40- & \cellcolor[HTML]{F8A102}40- & \cellcolor[HTML]{F56B00}50- & \cellcolor[HTML]{F8A102}40- \\ \hline
\multicolumn{1}{|c|}{\cellcolor[HTML]{EFEFEF}3}  & \cellcolor[HTML]{F8FF00}20- & \cellcolor[HTML]{F8FF00}20- & \cellcolor[HTML]{F8FF00}20- & \cellcolor[HTML]{F8FF00}20- & \cellcolor[HTML]{F8FF00}20- & \cellcolor[HTML]{F8FF00}20- & \cellcolor[HTML]{34FF34}10- & \cellcolor[HTML]{34FF34}10- & \cellcolor[HTML]{34FF34}10- & \cellcolor[HTML]{34FF34}10- & \cellcolor[HTML]{34FF34}10- & \cellcolor[HTML]{34FF34}10- \\ \hline
\multicolumn{1}{|c|}{\cellcolor[HTML]{EFEFEF}4}  & \cellcolor[HTML]{FFCC67}30- & \cellcolor[HTML]{FFCC67}30- & \cellcolor[HTML]{34FF34}10- & \cellcolor[HTML]{FFCC67}30- & \cellcolor[HTML]{FFCC67}30- & \cellcolor[HTML]{34FF34}10- & \cellcolor[HTML]{F8FF00}20- & \cellcolor[HTML]{FFCC67}30- & \cellcolor[HTML]{34FF34}10- & \cellcolor[HTML]{F8FF00}20- & \cellcolor[HTML]{FFCC67}30- & \cellcolor[HTML]{F8FF00}20- \\ \hline
\multicolumn{1}{|c|}{\cellcolor[HTML]{EFEFEF}5}  & \cellcolor[HTML]{34FF34}10- & \cellcolor[HTML]{34FF34}10- & \cellcolor[HTML]{34FF34}10- & \cellcolor[HTML]{34FF34}10- & \cellcolor[HTML]{34FF34}10- & \cellcolor[HTML]{34FF34}10- & \cellcolor[HTML]{34FF34}10- & \cellcolor[HTML]{34FF34}10- & \cellcolor[HTML]{34FF34}10- & \cellcolor[HTML]{34FF34}10- & \cellcolor[HTML]{34FF34}10- & \cellcolor[HTML]{F8FF00}20- \\ \hline
\multicolumn{1}{|c|}{\cellcolor[HTML]{EFEFEF}\textbf{6}}  & \cellcolor[HTML]{34FF34}10- & \cellcolor[HTML]{34FF34}10- & \cellcolor[HTML]{34FF34}10- & \cellcolor[HTML]{34FF34}10- & \cellcolor[HTML]{34FF34}10- & \cellcolor[HTML]{34FF34}10- & \cellcolor[HTML]{34FF34}10- & \cellcolor[HTML]{34FF34}10- & \cellcolor[HTML]{34FF34}10- & \cellcolor[HTML]{34FF34}10- & \cellcolor[HTML]{34FF34}10- & \cellcolor[HTML]{F8FF00}20- \\ \hline
\multicolumn{1}{|c|}{\cellcolor[HTML]{EFEFEF}7}  & \cellcolor[HTML]{34FF34}10- & \cellcolor[HTML]{34FF34}10- & \cellcolor[HTML]{34FF34}10- & \cellcolor[HTML]{34FF34}10- & \cellcolor[HTML]{34FF34}10- & \cellcolor[HTML]{34FF34}10- & \cellcolor[HTML]{34FF34}10- & \cellcolor[HTML]{34FF34}10- & \cellcolor[HTML]{34FF34}10- & \cellcolor[HTML]{34FF34}10- & \cellcolor[HTML]{34FF34}10- & \cellcolor[HTML]{34FF34}10- \\ \hline
\multicolumn{1}{|c|}{\cellcolor[HTML]{EFEFEF}8}  & \cellcolor[HTML]{34FF34}10- & \cellcolor[HTML]{34FF34}10- & \cellcolor[HTML]{34FF34}10- & \cellcolor[HTML]{34FF34}10- & \cellcolor[HTML]{34FF34}10- & \cellcolor[HTML]{34FF34}10- & \cellcolor[HTML]{34FF34}10- & \cellcolor[HTML]{34FF34}10- & \cellcolor[HTML]{34FF34}10- & \cellcolor[HTML]{34FF34}10- & \cellcolor[HTML]{34FF34}10- & \cellcolor[HTML]{34FF34}10- \\ \hline
\multicolumn{1}{|c|}{\cellcolor[HTML]{EFEFEF}9}  & \cellcolor[HTML]{34FF34}10- & \cellcolor[HTML]{34FF34}10- & \cellcolor[HTML]{34FF34}10- & \cellcolor[HTML]{34FF34}10- & \cellcolor[HTML]{34FF34}10- & \cellcolor[HTML]{34FF34}10- & \cellcolor[HTML]{34FF34}10- & \cellcolor[HTML]{34FF34}10- & \cellcolor[HTML]{34FF34}10- & \cellcolor[HTML]{34FF34}10- & \cellcolor[HTML]{34FF34}10- & \cellcolor[HTML]{34FF34}10- \\ \hline
\multicolumn{1}{|c|}{\cellcolor[HTML]{EFEFEF}10} & \cellcolor[HTML]{34FF34}10- & \cellcolor[HTML]{34FF34}10- & \cellcolor[HTML]{34FF34}10- & \cellcolor[HTML]{34FF34}10- & \cellcolor[HTML]{34FF34}10- & \cellcolor[HTML]{34FF34}10- & \cellcolor[HTML]{34FF34}10- & \cellcolor[HTML]{34FF34}10- & \cellcolor[HTML]{34FF34}10- & \cellcolor[HTML]{34FF34}10- & \cellcolor[HTML]{34FF34}10- & \cellcolor[HTML]{34FF34}10- \\ \hline
\end{tabular}
}%
\vspace{0.2em}
\resizebox{\columnwidth}{!}{
\begin{tabular}{|l|
>{\columncolor[HTML]{34FF34}}l |l|l|
>{\columncolor[HTML]{F8FF00}}l |l|l|
>{\columncolor[HTML]{FFCC67}}l |}
\cline{1-2} \cline{4-5} \cline{7-8}
$< 10\%$ from min & 10- & & $< 20\%$ from min & 20- & & $< 30\%$ from min & 30- \\ \cline{1-2} \cline{4-5} \cline{7-8}
\end{tabular}
}%
\vspace{0.1em}
\resizebox{\columnwidth}{!}{
\begin{tabular}{|l|
>{\columncolor[HTML]{F8A102}}l |l|l|
>{\columncolor[HTML]{F56B00}}l |l|l|
>{\columncolor[HTML]{FE0000}}l |}
\cline{1-2} \cline{4-5} \cline{7-8}
$< 40\%$ from min & 40- & & $< 50\%$ from min & 50- & & $\ge 50\%$ from min & 50+ \\ \cline{1-2} \cline{4-5} \cline{7-8}
\end{tabular}
}%
\vspace{0.2em}
\caption{Evaluation of various weights for tree selection ($\mathcal{F}$, $\mathcal{S}$ and $\mathcal{M}$ refer to scheduling policies FCFS, SRPT and Fair Sharing, respectively)}\label{fig:weights.assignment}
\end{figure}

Figure \ref{fig:weights.assignment} shows our simulation results of receiver completion times for bulk multicast transfers with $10$ receivers for a fixed arrival rate of $\lambda=1$. We considered both light-tailed and heavy-tailed transfer volume distributions. Techniques \#1, \#7, \#8, \#9 and \#10 all used minimal edge Steiner trees, and so offer minimum bandwidth usage. However, this comes at the cost of increasing completion times especially when edges have a non-homogeneous capacity. Techniques \#2 and \#4 use utilization as criteria for load balancing. Minimizing maximum link utilization has long been a popular objective for traffic engineering over WAN. As can be seen, they have the highest bandwidth usage compared to other techniques (up to $40\%$ above the minimum) for almost all scenarios while their completion times are at least $20\%$ worse than the minimum for several scenarios. Techniques \#3, \#5, and \#6 operate based on link load (i.e., total outstanding volume of traffic per edge) among which technique \#3 (minimizing maximum load) has the highest variation between best and worst case performance (up to $40\%$ worse than the minimum in mean completion times). Techniques \#5 and \#6 (minimizing the sum of load including and excluding the new multicast request) on the other hand offer consistently good performance that is up to $13\%$ above the minimum (for all performance metrics) across all scheduling policies, topologies, and traffic patterns. These techniques offer lower completion times for the GEANT topology with non-uniform link capacity. Technique \#6 also provides slightly better bandwidth usage and better completion times compared to \#5 for the majority of scenarios (not shown). Our proposals rely on technique \#6 for selection of load-aware forwarding trees, as shown in Algorithm \ref{algo_dccast}.

\subsection{Receiver Set Partitioning}
Receiver set partitioning allows separation of faster receivers from the slowest (or slower ones). This is essential to improve network utilization and speed up transfers when there are competing transfers or physical bottlenecks. For example, both GEANT and UNINETT have edges that vary by at least a factor of $10\times$ in capacity. We evaluate QuickCast over a variety of scenarios.

\subsubsection{Effect of Number of Receivers}
\begin{figure*}[t]
    \centering
    \subfigure[$\lambda = 0.001$ (light load)]
    {
        \includegraphics[width=\textwidth]{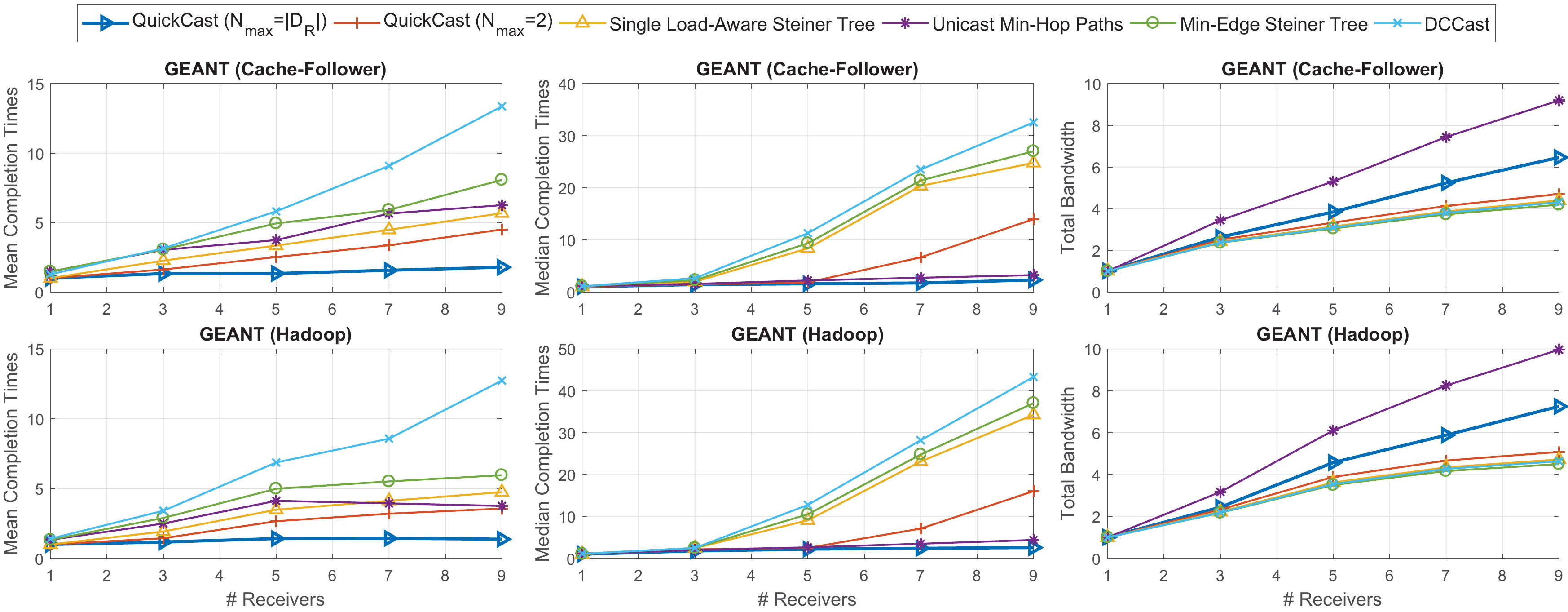}
    }
    \\
    \subfigure[$\lambda = 1$ (heavy load)]
    {
        \includegraphics[width=\textwidth]{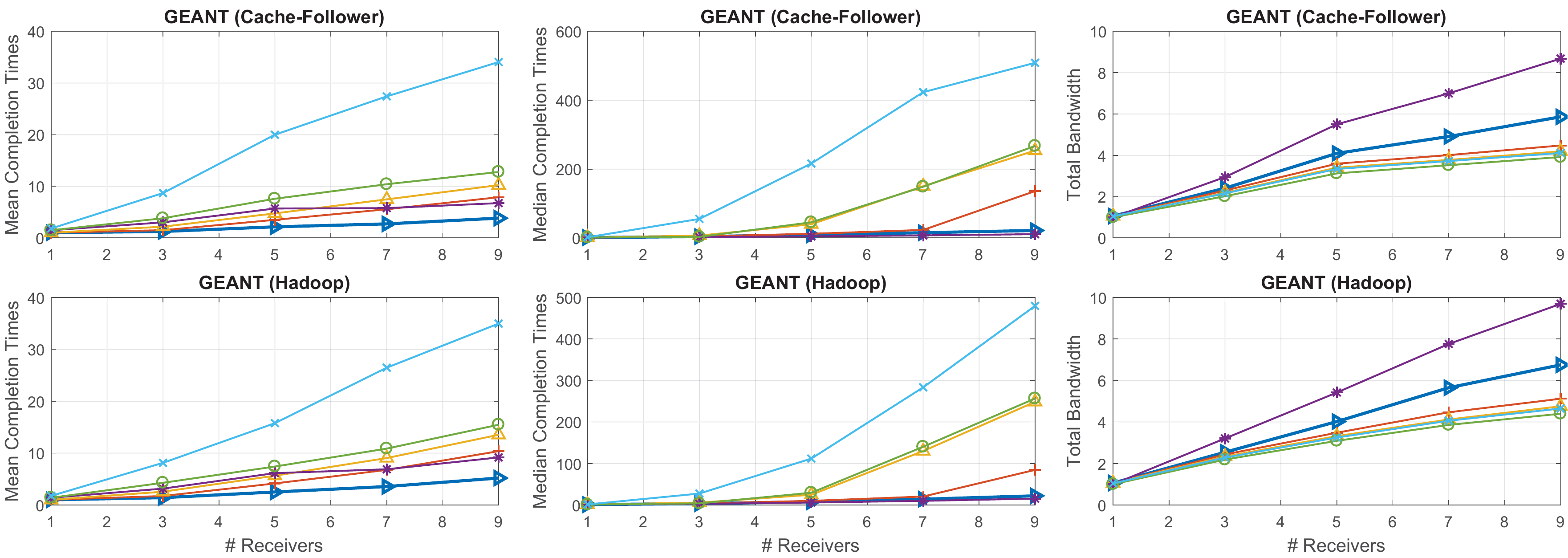}
    }
    \caption{Various schemes for bulk multicast transfers. All schemes use max-min fair rates except for DCCast which uses FCFS and are performed on GEANT topology. Plots are normalized by minimum (lower is better). We used Cache-Follower and Hadoop traffic patterns in Table \ref{table_traffic}.}
    \label{fig:overall}
\end{figure*}

We provide an overall comparison of several schemes (QuickCast, Single Load-Aware Steiner Tree, and DCCast \cite{dccast}) along with two basic solutions of using a minimum edge Steiner tree and unicast minimum hop path routing as shown in Figure \ref{fig:overall}. We also considered both light and heavy load regimes. We used real inter-datacenter traffic patterns reported by Facebook for two applications of Cache-Follower and Hadoop \cite{social_inside}. Also, all schemes use the fair sharing rate allocation based on max-min fairness except DCCast which uses the FCFS policy.

The minimum edge Steiner tree leads to the minimum bandwidth consumption. The unicast minimum hop path routing approach separates all receivers per bulk multicast transfer. It, however, uses a significantly larger volume of traffic and also does not offer the best mean completion times for the following reasons. First, it exhausts network capacity quickly which increases tail completion times by a significant factor (not shown here). Second, it can lead to many additional shared links that increase contention across flows and reduce throughput per receiver. The significant increase in completion times of higher percentiles increases the average completion times of the unicast approach.

With $N_{max}=\lvert \pmb{\mathrm{D}}_R \rvert$, we see that QuickCast offers the best mean and median completion times, i.e., up to $2.84\times$ less compared to QuickCast with $N_{max}=2$, up to $3.64\times$ less compared to unicast minimum hop routing, and up to $3.33\times$ less than single load-aware Steiner tree. To achieve this gain, QuickCast with $N_{max}=\lvert \pmb{\mathrm{D}}_R \rvert$ uses at most $1.49\times$ more bandwidth compared to using minimum edge Steiner trees which is still $1.4\times$ less than bandwidth usage of unicast minimum hop routing. We also see that while increasing the number of receivers, QuickCast with $N_{max}=\lvert \pmb{\mathrm{D}}_R \rvert$ offers consistently small median completion times by separating fast and slow receivers since the number of partitions are not limited. Overall, we see a higher gain under light load as there is more capacity available to utilize. We also recognize that QuickCast with either $N_{max}=2$ or $N_{max}=\lvert \pmb{\mathrm{D}}_R \rvert$ performs almost always better than unicast minimum hop routing in mean completion times.

\subsubsection{Speedup by Receiver Rank}
\begin{figure*}[t]
    \centering
    \includegraphics[width=\textwidth]{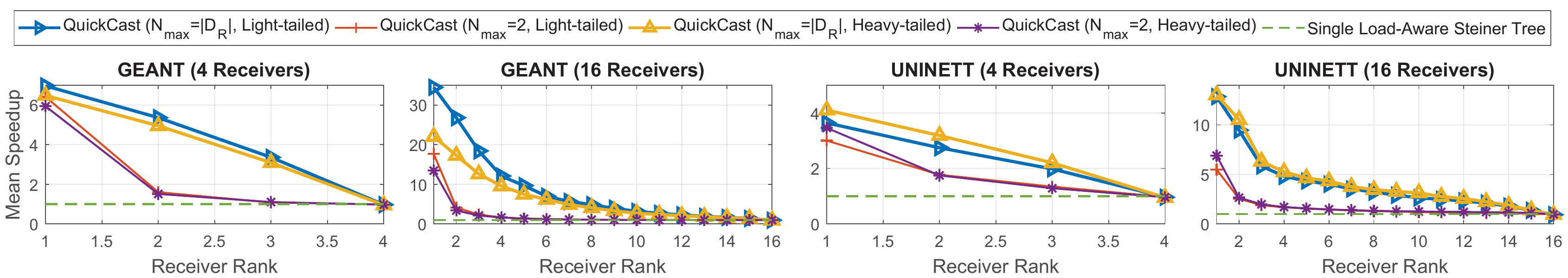}
    \caption{Mean receiver completion time speedup (larger is better) of receivers compared to single load-aware Steiner tree (Algorithm \ref{algo_dccast}) by their rank (receivers sorted by their speed from fastest to slowest per transfer), receivers selected according to uniform distribution from all nodes, we considered $\lambda = 1$.}
    \label{fig:speedup_1}
\end{figure*}

Figure \ref{fig:speedup_1} shows how QuickCast can speed up multiple receivers per transfer by separating them from the slower receivers. The gains are normalized by when a single partition is used per bulk multicast transfer. In case the number of partitions is limited to two similar to \cite{quickcast}, the highest gain is usually obtained by the first two to three receivers while allowing more partitions, we can get considerably higher gain for a significant fraction of receivers. Also, by not limiting the partitions to two, we see higher gains for all receiver ranks that is above $2\times$ for multiple receiver ranks. This comes at the cost of higher bandwidth consumption which we saw earlier in the previous experiment.

\subsubsection{Partitioning Factor ($p_f$)}
\begin{figure*}[t]
    \centering
    \includegraphics[width=\textwidth]{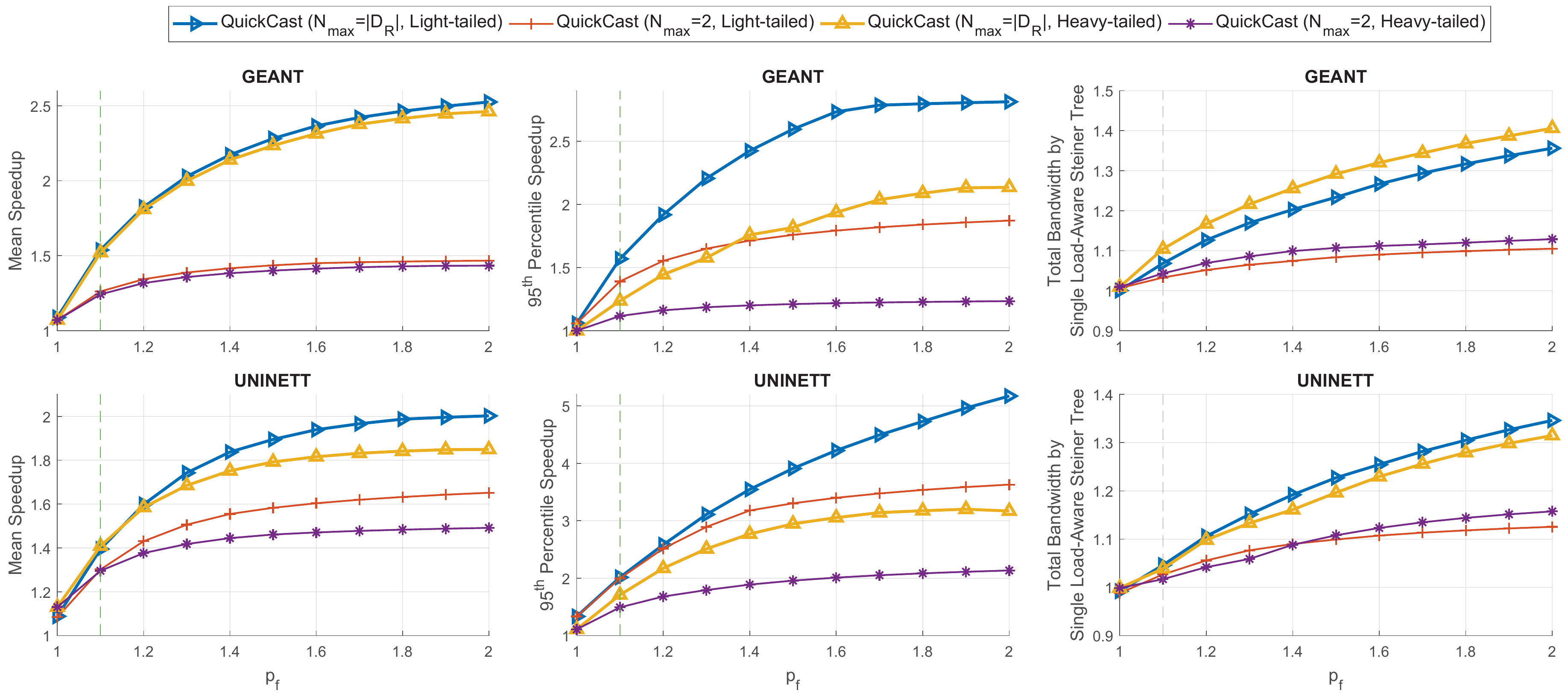}
    \caption{Performance of QuickCast as a function of partitioning factor $p_f$. We assumed 4 receivers and an arrival rate of $\lambda = 1$.}
    \label{fig:pf}
\end{figure*}

The performance of QuickCast as a function of the partitioning factor has been shown in Figure \ref{fig:pf} where gains are normalized by single load-aware Steiner tree which uses a single partition per bulk multicast transfer. We computed per receiver mean and 95\textsuperscript{th} percentile completion times as well as bandwidth usage. As can be seen, bandwidth consumption increases with partitioning factor as more requests' receivers are partitioned into two or more groups. The gains in completion times keep increasing if $N_{max}$ is not limited as we increase $p_f$. That, however, can ultimately lead to unicast delivery to all receivers (every receiver as a separate partition) and excessive bandwidth usage. We see a diminishing return type of curve as $p_f$ is increased with the highest returns coming when we increase $p_f$ from 1 to 1.1 (marked with a green dashed lined). That is because using too many partitions can saturate network capacity while not improving the separation of fast and slow nodes considerably. At $p_f=1.1$, we see up to 10\% additional bandwidth usage compared to single load-aware Steiner tree while mean completion times improve by between 40\% to 50\%. According to other experiments not shown here, with large $p_f$, it is possible even to see reductions in gain that come from excessive bandwidth consumption and increased contention over capacity. Note that this experiment was performed considering four receivers per bulk multicast transfer. Using more receivers can lead to more bandwidth usage with the same $p_f$, an increased slope at values of $p_f$ close to 1, and faster saturation of network capacity as we increase $p_f$. Therefore, using smaller $p_f$ is preferred with more receivers per transfer.

\begin{figure*}[t]
    \centering
    \includegraphics[width=0.9\textwidth]{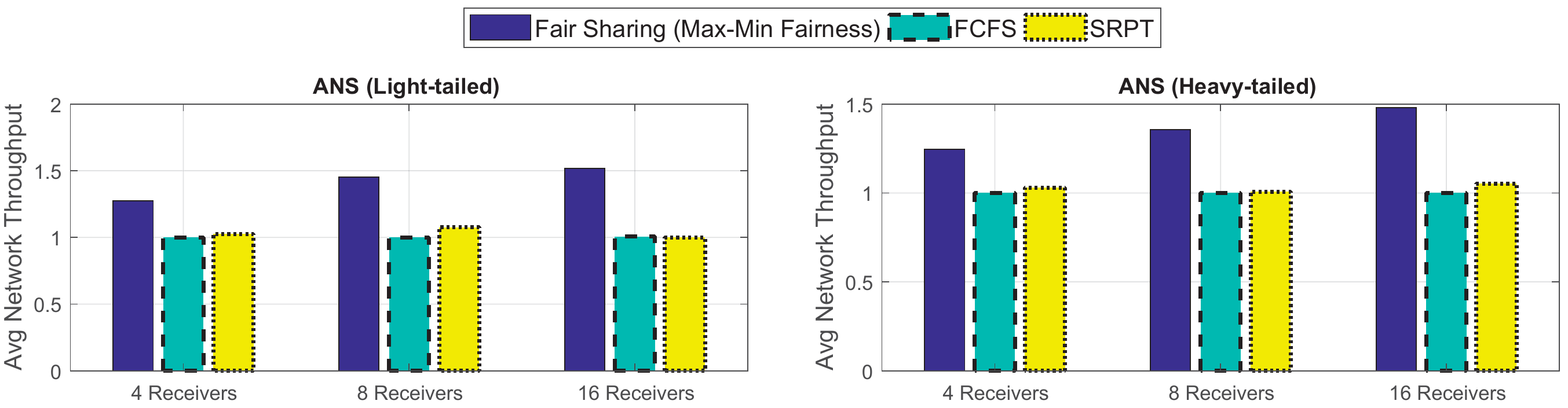}
    \caption{Average throughput of bulk multicast transfers obtained by running different scheduling policies. We started 100 transfers at the time zero, senders and receivers were selected according to the uniform distribution. Each group of bars is normalized by the minimum in that group.}
    \label{fig:thr}
\end{figure*}

\subsection{Effect of Rate Allocation Policies} \label{eval-rate-alloc}
As explained earlier in \S \ref{rate-allocation}, when scheduling traffic over large forwarding trees, fair sharing can sometimes offer significantly higher throughput and hence better completion times. We performed an experiment over the ANS topology and with both light-tailed and heavy-tailed traffic distributions. ANS topology has uniform link capacity across all edges which helps us rule out the effect of capacity variations on throughput obtained via different scheduling policies. We also considered an increasing number of receivers from 4 to 8 and 16. Figure \ref{fig:thr} shows the results. We see that fair sharing offers a higher average throughput across all ongoing transfers compared to FCFS and SRPT and that with more receivers, the benefit of using fair sharing increases to up to $1.5\times$ with 16 receivers per transfer.

\subsection{Running Time}
To ensure scalability of proposed algorithms, we measured the running time of our algorithms over various topologies (with different sizes) and with varying rates of arrival. We assumed two arrival rates of $\lambda=0.001$ and $\lambda=1$ which account for light and heavy load regimes. We also considered eight receivers per transfer and all the three topologies of ANS, GEANT, and UNINETT. We saw that the running time of Algorithm \ref{algo_dccast}, and \ref{algo_quick} remained below one millisecond and 20 milliseconds, respectively, across all of these scenarios. These numbers are less than the propagation latency between the majority of senders and receivers over considered topologies (a simple TCP handshake would take at least twice the propagation latency). More efficient realization of these algorithms can further reduce their running time (e.g., implementation in C/C++ instead of Java).

\subsection{Forwarding Plane Resource Usage}
QuickCast can be realized using software-defined networking and OpenFlow compatible switches. To forward packets to multiple outgoing ports on switches where trees branch out to numerous edges, we can use group tables which have been supported by OpenFlow since early versions. Besides, an increasing number of physical switch makers have added support for group tables. To allow forwarding to multiple outgoing ports, the group table entries should be of type ``ALL", i.e., \texttt{OFPGT\_ALL} in the OpenFlow specifications. Group tables are highly scarce (compared to TCAM entries) and so should be used with care. Some new switches support 512 or 1024 entries per switch. Another critical parameter is the maximum number of action buckets per entry which primarily determines the maximum possible branching degree for trees. Across the switches we looked at, we found that the minimum supported value was 8 action buckets which should be enough for WAN topologies as most of such do not have any nodes with this connectivity degree.

In general, reasoning about the number of group table entries needed to realize different schemes is hard since it depends on how the trees are formed which is highly intertwined with edge weights that depend on the distribution of load. For example, consider a complete binary tree with 8 receivers as leaves and the sender at the root. This will require 6 group table entries to transmit to all receivers with two action buckets per each intermediate node on the tree (branching at the sender does not need a group table entry). If instead, we used an intermediate node to connect to all receivers with a branching degree of 8, we would only need one group table entry with eight action buckets.

We measured the number of group table entries needed to realize QuickCast. We computed the average of the maximum, and maximum of the maximum number of entries used per switch during the simulation for the topologies of ANS, GEANT, and UNINETT, with arrival rates of $\lambda=0.001$ and $\lambda=1$, considering both light-tailed and heavy-tailed traffic patterns and assuming that each bulk multicast transfer had eight receivers. The experiment was terminated when 200 transfers arrived. Looking at the maximum helps us see whether there are enough entries at all times to handle all concurrent transfers. Interestingly, we saw that by using multiple trees per transfer, both the average and maximum of the maximum number of group table entries used were less than when a single tree was used per transfer. One reason is that using a single tree slows down faster receivers which may lead to more concurrent receivers that increase the number of group entries. Also, by partitioning receivers, we make subsequent trees smaller and allow them to branch out closer to their receivers which balances the use of group table entries usage across the switches reducing the maximum. Finally, by using more partitions, the maximum number of times a tree needs to branch to reach all of its receivers decreases. Across all the scenarios considered above, the maximum of maximum group table entries at any timeslot was 123, and the average of the maximum was at most 68 for QuickCast. Furthermore, by setting $N_{max}=\lvert \pmb{\mathrm{D}}_R \rvert$ which allows for more partitions, the maximum of maximum group table entries decreased by up to 17\% across all scenarios.

\section{Conclusions and Future Work}
A variety of applications running across datacenters replicate content between geographically dispersed sites for increased availability and reliability. Such data replication generates bulk multicast transfers with an apriori known sender and set of receivers per transfer which can be efficiently performed using multicast forwarding trees. We introduced the bulk multicast routing and scheduling problem with the objective of minimizing mean completion times of receivers and decomposed it into three sub-problems of receiver set partitioning, tree selection, and rate allocation. We then proposed QuickCast which applies three heuristic techniques to offer approximate solutions to these three hard sub-problems. For future research, we will consider parallel trees to increase throughput which is especially helpful under light traffic load. Also, application of BIER, which allows dynamic and low-cost updates to the forwarding trees, opens up new research opportunities.

\bibliographystyle{IEEEtran}
\bibliography{citations.bib}

\end{document}